\documentclass[12pt]{article}
\pdfoutput =1
\usepackage{float} 
\usepackage{amsmath}
\usepackage{amsfonts}   
\usepackage{amssymb}

\begin{document}
\date{}
\title{{\bf{\Large Hall viscosity to entropy ratio in higher derivative theories}}}
\author{
 {\bf {\normalsize Dibakar Roychowdhury}$
$\thanks{E-mail:  dibakarphys@gmail.com, dibakar@cts.iisc.ernet.in}}\\
 {\normalsize Centre for High Energy Physics, Indian Institute of Science, }
\\{\normalsize C.V. Raman Avenue, Bangalore 560012, Karnataka, India}
}

\maketitle
\begin{abstract}
In this paper based on the basic principles of gauge/gravity duality we compute the hall viscosity to entropy ratio in the presence of various higher derivative corrections to the dual gravitational description embedded in an asymptotically $ AdS_{4} $ space time. As the first step of our analysis, considering the back reaction we impose higher derivative corrections to the abelian gauge sector of the theory where we notice that the ratio indeed gets corrected at the leading order in the coupling. Considering the probe limit as a special case we compute this leading order correction over the fixed background of the charged black brane solution. Finally we  consider higher derivative ($ R^{2} $) correction to the gravity sector of the theory where we notice that the above ratio might get corrected at the sixth derivative level.
 \end{abstract}

\section{Overview and Motivation}
In the recent years the AdS/CFT correspondence \cite{ref1}-\cite{ref3} has been found to provide an extremely elegant set up to study various strongly coupled phenomena (at finite temperature or at finite chemical potential) that take place either in the usual condensed matter systems \cite{ref4}-\cite{ref5} or in a relativistic fluid \cite{Policastro:2001yc}-\cite{Kovtun:2004de}. The study of the relativistic hydrodynamics using holographic frameworks has attracted lots of attention due to its various remarkable features, for example using the holographic techniques one can in fact compute the hydrodynamic transport coefficients for a relativistic strongly coupled viscous fluid at finite temperature. All such analysis are based on the crucial fact that the long wavelength dynamics of a strongly coupled plasma at the boundary of an asymptotically AdS space time could be described in terms of the near horizon data of the black brane solution in the bulk  space time.

One of the major successes of the AdS/CFT correspondence is that it provides an universal lower bound for the shear viscosity to entropy ratio ($ \eta_{s}/s $) for a wide class of gauge theories that admit a dual gravitational description which is basically the classical two derivative Einstein gravity in an asymptotically Anti-de-Sitter (AdS) space \cite{Saremi:2006ep}-\cite{Cai:2008ph}. However it has been observed that in certain higher derivative theories of gravity this bound is not respected \cite{Brigante:2007nu}-\cite{Banerjee:2009wg}. As a matter of fact this has been confirmed that the class of gauge theories that respect the shear viscosity bound must have their dual counterpart as the Einstein's theory of gravity\footnote{Any non trivial interaction between matter fields does not seem to affect this bound.} \cite{Brustein:2008cg}. This claim was further strengthened in certain holographic calculations where there was an attempt to study whether any higher derivative corrections imposed solely on the (abelian) gauge sector of the Einstein Maxwell gravity could indeed affect the universality of the lower bound of the $ \eta_{s}/s $ ratio \cite{Cai:2008in}-\cite{Tan:2009yg}.

Besides being an extremely elegant tool to compute the hydrodynamic transport coefficients at strong coupling, the gauge/gravity duality also provides a precise frame work to study hydrodynamic systems in the presence of parity violating effects in ($ 2+1 $) dimensions \cite{Saremi:2011ab}-\cite{Delsate:2011qp}. It is well known from the knowledge of standard hydrodynamics that the systems with parity violation contain additional response parameters/transport coefficients at the long wavelength limit. One of the well studied example of such a transport coefficient is the coefficient of hall viscosity ($ \eta_A $) that appears as the transport coefficient in the first order viscous hydrodynamics in ($ 2+1 $) dimensions. It is noteworthy to mention that, like the other transport coefficients it is also extremely difficult to compute the hall conductivity coefficient ($ \eta_A $) at strong coupling. However such a situation could be evaded once we consider the AdS/CFT frame work where considering a dual gravitational description it is indeed possible to compute the coefficient of hall viscosity ($ \eta_{A} $) at strong coupling. The hall viscosity to entropy ratio ($ \eta_{A}/s $) corresponding to a neutral black brane was first computed by Saremi and Son in \cite{Saremi:2011ab}. Later on considering the probe limit this analysis was further extended for the charged black branes \cite{Chen:2011fs} in the Einstein Maxwell theory of gravity. In both the cases it was observed that the ratio $ \eta_{A}/s $ possesses some unique structure.

 Keeping the spirit of the earlier discussions regarding the higher derivative corrections on the $ \eta_{s}/s $ ratio, the question that naturally arises here is whether the hall viscosity to entropy ratio ($ \eta_{A}/s $) is also modified in the presence of higher derivative corrections to the usual Einstein Maxwell theory of gravity.  Till date this issue has never been addressed in the literature. The purpose of the present article is to address this issue considering two specific examples. In the first example, we consider a particular holographic set up where we take in to account the higher derivative (Born-Infeld (BI)) corrections to the (abelian) gauge sector of the Einstein Maxwell theory of gravity\footnote{Our analysis is partly motivated from the previous studies \cite{Cai:2008in}-\cite{Tan:2009yg} where it was observed that in the presence of Born-Infeld (BI) corrections to the abelian gauge sector of the theory the conjectured bound on the shear viscosity to entropy ratio does not get affected. Following the same spirit it would be then quite interesting to see whether the $ \eta_{A}/s $ ratio gets affected in the presence of similar higher derivative corrections.} \cite{Cai:2008in}-\cite{Tan:2009yg}. At this stage it is customary to mention that from the present analysis one can easily trace back the corresponding result for the standard Einstein Maxwell theory \cite{Chen:2011fs} by setting the coupling constant associated with the higher derivative corrections equal to zero.  From our calculations it turns out that the coefficient of hall viscosity ($ \eta_{A} $) apart from having its standard form also contains some additional structure at the leading order in the higher derivative (BI) coupling.  Moreover we carry out the entire analysis considering the back reaction on the metric since the value of the ratio $ \eta_{A}/s  $ is not yet known for the back reacted geometry. As the final solution corresponding to the back reacted space time turned out to be too complicated to solve exactly, therefore considering the \textit{probe} limit, we compute the leading order correction to $ \eta_{A} $ over the fixed background of the charged black brane solution in the Einstein BI gravity.
 
 In the second example, we consider higher derivative ($ R^{2} $) corrections to the Einstein Hilbert sector of the theory in the presence of the gravitational Chern-Simons term that was originally incorporated in \cite{Saremi:2011ab}. From our calculations we note that $ \eta_{A}/s $ ratio does not receive any correction upto fourth order derivatives. On the other hand, from our computations we get a clear indication that $ \eta_{A}/s $ ratio might receive corrections at the sixth derivative level. 

The organisation of the paper is the following: In Section 2, to start with we first give a brief description of the holographic set up where we incorporate higher derivative (BI) corrections to the gauge sector of the Einstein Maxwell theory. Next we compute the hall viscosity associated with the stress tensor of the boundary hydrodynamics following a systematic derivative expansion \cite{Bhattacharyya:2008jc} in the bulk $ AdS_{4} $.  As per as the shear viscosity is concerned our result seems to be quite compatible to that with the earlier analysis \cite{Cai:2008in}-\cite{Tan:2009yg}. Finally, considering the probe limit we carry out an exercise where we actually compute the leading order correction to $ \eta_A $ over a fixed background. In Section 3, we discuss the effect of $ R^{2} $ corrections on $ \eta_A $. Finally, we conclude in Section 4.

\section{Boosted Born-Infeld black branes}
Before we actually start our analysis it is always good to have a brief overview of the computations that we are going to do. The basic idea is the following: Considering the \textit{back reaction}, as a first step we construct the boosted black brane solution in Einstein Born-Infeld gravity \cite{Tan:2009yg} in an asymptotically $ AdS_{4} $ space time. We also add a gravitational Chern-Simons (CS) term in our theory that introduces a parity breaking interaction in the bulk \cite{Saremi:2011ab}. To preserve the parity we associate a neutral pseudo scalar field ($ \theta $) with this gravitational CS term. Therefore the whole combination together preserves the parity in the bulk. Finally, in order to compute the response parameters associated with the first order viscous hydrodynamics at the boundary of the $ AdS_{4} $, we consider a derivative expansion\footnote{By derivative expansion we always mean derivatives with respect to the boundary coordinates ($ x^{\mu}, \mu=1,2,3 $) \cite{Bhattacharyya:2008jc}.} of the velocity field ($ u^{\mu} $), temperature ($ T $) as well as the charge ($ Q $) in the bulk space time and then solve the Einstein's equation perturbatively upto leading order in that derivative expansion.

\subsection{The holographic model}

We start with the following action in $ (3+1) $ dimensions namely,
\begin{eqnarray}
S = \int d^{4}x \sqrt{-g} [ \mathcal{L} + \mathcal{B}(F)]
\end{eqnarray}
where the Lagrangian is given by \cite{Saremi:2011ab},
\begin{eqnarray}
\mathcal{L}= R- 2\Lambda - \frac{1}{2}(\partial \theta)^{2}- V(\theta)- \frac{\lambda}{4} \theta\ ^{\ast}R R 
\end{eqnarray}
with the axion potential \cite{Chen:2011fs},
\begin{eqnarray}
V(\theta)= \frac{1}{2}m^{2}\theta^{2}+\frac{1}{4}c\theta^{4}.
\end{eqnarray}

Here $ \mathcal{B}(F) $ is the Born-Infeld (BI) Lagrangian that could be formally expressed as\footnote{Here $ b $ is the BI parameter. Note that in the limit $ b\rightarrow\infty $ one generally recovers the familiar Maxwell Lagrangian.}'\footnote{Note that studying the leading order effect in the BI coupling eventually implies that what we are studying is basically the effect of $ F^{4} $ corrections on the $ \eta_A/s $ ratio. Similar analysis have been made earlier in the context of $ \eta_s/s $ ratio in \cite{Cai:2008ph}, \cite{Banerjee:2009ju}.},
\begin{eqnarray}
\mathcal{B}(F)&=& 4b^{2}\left(1-\sqrt{1+\frac{F^{2}}{2b^{2}}} \right)\nonumber\\
&=& -F^{2}-\frac{F^{4}}{8b^{2}}+\mathcal{O}(1/b^{4}).
\end{eqnarray}

Note that here $ F^{2}=\frac{1}{4}F_{ab}F^{ab} $ and $ \Lambda (=-3) $ is the cosmological constant\footnote{Here we have set the AdS length scale to unity.}. The quantity $ ^{\ast}R R $ could be formally expressed as \cite{Saremi:2011ab},
\begin{eqnarray}
^{\ast}R R = \frac{1}{2} \frac{\epsilon^{abcd}}{\sqrt{-g}}R^{m}\ _{ncd}R^{n}\ _{mab}.
\end{eqnarray}

The natural next step would be to compute the equations of motion. Let us first note down the Einstein's equation as well as the Maxwell Born-Infeld equation,
\begin{eqnarray}
G_{ab}+ g_{ab}\Lambda - \lambda C_{ab}&=& T_{ab}(\theta)+T_{ab}(F)\nonumber\\
\nabla_{a}\left( \frac{F^{ab}}{\sqrt{1+\frac{F^{2}}{2b^{2}}}}\right)&=&0\label{E1}
\end{eqnarray}
along with the scalar equation of motion,
\begin{eqnarray}
\nabla^{2}\theta &=&\frac{\partial V}{\partial \theta}+ \frac{\lambda}{4}\ ^{\ast}R R\label{E2}
\end{eqnarray}
where $ C_{ab} $ is the symmetric (traceless) Cotton tensor and $ T_{ab} $ is the energy momentum tensor that could be expressed as,
\begin{eqnarray}
C_{ab}&=& \nabla_{m}\nabla_n (\theta\ \ast R^{m}\ _{(ab)}\ ^{n})\nonumber\\
T_{ab}(\theta)&=& \frac{1}{2}\nabla_{a}\theta \nabla_{b}\theta -\frac{1}{4}g_{ab}(\partial \theta)^{2}-\frac{1}{2}g_{ab}V(\theta)\nonumber\\
T_{ab}(F)&=&\frac{1}{2}g_{ab}\mathcal{B}(F)+\frac{2F_{ac}F_{b}\ ^{c}}{\sqrt{1+\frac{F^{2}}{2b^{2}}}}.\label{e6}
\end{eqnarray}

\subsection{Fluid/gravity approach}

In this section our goal would be to solve Einstein's equation order by order in the perturbation series. To do that we shall first write down the metric expanded upto first order in the boundary derivatives about the origin $ x^{\mu}=0 $. Once the solution (metric) is obtained around the origin it could be further extended for the rest of the manifold iteratively \cite{Bhattacharyya:2008jc}.

Let us first note down the ansatz that satisfies the above equations of motion (\ref{E1})  and (\ref{E2}) namely\footnote{Since the background possesses the spatial $ SO(2) $ symmetry therefore one can in principle solve the scalar, vector and tensor equations separately. However for the present case of study it will be sufficient for us to consider the tensor equation only.},
\begin{eqnarray}
ds^{2}&=&- 2H(r,T,Q)u_{\mu}dx^{\mu}dr-r^{2}F(r,T,Q)u_{\mu}u_{\nu}dx^{\mu}dx^{\nu}+r^{2}(\eta_{\mu\nu}+u_{\mu}u_{\nu})dx^{\mu}dx^{\nu}\nonumber\\
\theta &=&\theta (r,T,Q)\nonumber\\
A &=& A(r,T,Q)u_{\mu}dx^{\mu}\label{E3}
\end{eqnarray}
where $ T $ is the Hawking temperature and $ Q $ is the charge of the black brane.

The above ansatz (\ref{E3}) eventually describes a charged boosted black brane solution that describes the boundary hydrodynamics in a derivative expansion along the boundary coordinates in ($ 2+1 $) dimensions. Here $ u^{\mu} $ is the three velocity which may be expressed as\footnote{$u^{\mu}$ is normalized such that $u^{\mu}u_{\mu}=-1$.},
\begin{eqnarray}
u^{\mu}=\frac{1}{\sqrt{1-\beta^{2}}}(1, \beta^{i}).
\end{eqnarray}

The entire analysis is based on the two major facts, first one is that we shall lift the constant entities like $ T $, $ Q $ and $ u^{\mu} $ to become slowly varying functions of the boundary coordinates ($ x^{\mu} $) and the second one is that we shall carry out the analysis in a co-moving frame where the fluid two velocity is zero at the origin of the boundary coordinates namely $ x^{\mu}=0 $. This enables us to write down the following expansion\footnote{Note that here $ i(=x,y) $ corresponds to the spatial directions at the boundary of the $ AdS_{4} $. Since the boundary is flat therefore the raising and the lowering of the spatial indices eventually does not matter.},
\begin{eqnarray}
u_{i}=u_{i}|_{x=0}+x^{\mu}\partial_{\mu}\beta_{i}=x^{\mu}\partial_{\mu}\beta_{i}.
\end{eqnarray}

Taylor expanding all the relevant thermodynamic entities upto first order in derivatives yields,
\begin{eqnarray}
ds^{2}&=&ds^{(0)2}+\varepsilon \left[2\Delta H dvdr - 2 H(r)x^{\mu}\partial_{\mu}\beta_{i}dx^{i}dr - r^{2}\Delta F dv^{2}+ 2r^{2}( F(r)-1) x^{\mu}\partial_{\mu}\beta_{i}dv dx^{i} \right]\nonumber\\
\theta &=& \theta (r)+ \varepsilon\Delta \theta\nonumber\\
A&=& -A(r)dv +\varepsilon [A(r)x^{\mu}\partial_{\mu}\beta_{i}dx^{i}-\Delta A dv]\label{E4} 
\end{eqnarray}
where the power of $ \varepsilon $ denotes the number of the derivatives. Here $ ds^{(0)2} $ is the metric corresponding to the zeroth order solution of Einstein's equations which is given by\footnote{This is the metric in the ingoing Eddington-Finkelstein coordinates.},
\begin{eqnarray}
ds^{(0)2}= 2 H(r)dv\ dr - r^{2} F(r)dv^{2}+r^{2}(dx^{2}+dy^{2}).
\end{eqnarray}

The general argument of the fluid/gravity approach is that one needs to add corrections to (\ref{E4}) order by order in $ \varepsilon $ so that the metric ($ g_{ab} $), the gauge field ($ A_{a} $) and the pseudo scalar field ($ \theta $) satisfy Eqs (\ref{E1}) and (\ref{E2}) order by order in the perturbation series.
In the following we note down the relevant corrections as,
\begin{eqnarray}
ds^{(1)2}_{corr}&=&\varepsilon \left(\frac{k(r)}{r^{2}}dv^{2}+2p(r)dvdr-r^{2}p(r)dx^{2}_{i}+\frac{2}{r}w_{i}(r)dvdx^{i} + 2 r^{2}\alpha_{xy}(r)dxdy\right)\nonumber\\
\theta_{corr} &=&\varepsilon \Theta\nonumber\\
A_{corr}&=& \varepsilon (a_{v}(r)dv + a_{i}(r)dx^{i}).\label{E5} 
\end{eqnarray} 
It is the combined form of (\ref{E4}) and (\ref{E5}) that satisfy Eq.(\ref{E1}) and (\ref{E2}) upto first order in the neighbourhood of $ x^{\mu}=0 $.

\subsection{Zeroth order computation}
Our goal in this section is to find an explicit analytical solution for the zeroth order ($ \mathcal{O}(\varepsilon^{0}) $) solutions ($ F(r) $ and $ H(r)$) of the (trace reversed) Einstein equations perturbatively upto leading order in the BI coupling ($ 1/b^{2} $). It is to be noted that throughout the analysis we shall take in to account the back reaction of the matter field ($ \theta $) on the metric ($ g_{ab} $) itself. Once these zeroth order solutions are known exactly, then one can proceed further to consider the first order metric fluctuations ($ \mathcal{O}(\varepsilon) $) in (\ref{E4}) and (\ref{E5}) in order to compute the hall viscosity coefficient ($ \eta_{A} $). It is noteworthy to mention that the zeroth order solutions obtained in this section are indeed going to affect the coefficient of the hall viscosity ($ \eta_{A} $) at the leading order ($ \mathcal{O}(\varepsilon) $). 

We shall start our analysis with the trace reversed form of the Einstein's equation (\ref{E1}) that turns out to be,
\begin{eqnarray}
R_{ab}+3g_{ab}-\lambda C_{ab}=T_{ab}(\theta)+T_{ab}(F)\label{E6}
\end{eqnarray}
where,
\begin{eqnarray}
T_{ab}(\theta)&=&\frac{1}{2}\left(\nabla_a \theta \nabla_b \theta +g_{ab}V(\theta) \right) \nonumber\\
T_{ab}(F)&=&\frac{1}{2}\left( F_{ac}F_{b}\ ^{c}-g_{ab}F^{2}\right) +\frac{3F^{2}}{16b^{2}}\left(g_{ab}F^{2}-\frac{2}{3}F_{ac}F_{b}\ ^{c} \right)+ \mathcal{O}(1/b^{4})\nonumber\\
&=& T^{(0)}_{ab}(F)+\frac{1}{b^{2}}t_{ab}(F) + \mathcal{O}(1/b^{4}).
\end{eqnarray}

Let us first note down the Einstein's equations (\ref{E6}) along with the gauge field (\ref{E1}) as well as the scalar equation (\ref{E2}) corresponding to the zeroth order solutions. These could be enumerated as follows, 
\begin{eqnarray}
F''-\frac{F'H'}{H}+\frac{6F'}{r}-\frac{2H'F}{rH}+\frac{6F}{r^{2}}-\frac{6H^{2}}{r^{2}}+\frac{H^{2}}{r^{2}}V(\theta)-\frac{A'^{2}}{2r^{2}}-\frac{A'^{4}}{32b^{2}r^{2}H^{2}}&=&0\nonumber\\
F'(r)-\frac{F}{H}H'+\frac{3F}{r}-\frac{3H^{2}}{r}+\frac{H^{2}}{2r}V(\theta)+\frac{A'^{2}}{4r}+\frac{3A'^{4}}{64b^{2}rH^{2}}&=&0\nonumber\\
\frac{H'}{H}-\frac{r}{4}\theta'^{2}&=&0\nonumber\\
A''\left(1+\frac{A'^{2}}{4b^{2}H^{2}}\right)+\left(\frac{2}{r}-\frac{H'}{H} \right) A'-\frac{A'^{3}H'}{4b^{2}H^{3}}&=&0 \nonumber\\
\theta'' +\theta' \left(\frac{4}{r}+\frac{F'}{F}-\frac{H'}{H} \right)-\frac{H^{2}}{r^{2}F}\frac{\partial V}{\partial \theta}&=&0 .\label{E7}
\end{eqnarray}

Eq.(\ref{E7}) basically represents a set of coupled non linear differential equations which have been expressed upto leading order in the BI coupling ($ 1/b^{2} $). In general it is indeed quite difficult to solve these equations exactly. Although using the perturbative techniques one can further split them in to different set of equations at each order in the perturbation series.
Quite naturally for us the next step would be to solve these equations (\ref{E7}) order by order in a perturbation series in the BI coupling ($ 1/b^{2} $). At this stage it is quite intuitive to note that we have five equations in hand and there are four variables to determine. Therefore in principle the system is solvable. 

After some trivial algebra one can note that the differential equations in (\ref{E7}) could be further simplified as,
\begin{eqnarray}
F'-\mathcal{A}(r)F+\mathcal{Q}(r)+\mathcal{G}(r,b^{2})&=&0\nonumber\\
A''\left(1+\frac{A'^{2}}{4b^{2}H^{2}}\right)+\left(\frac{2}{r}-\frac{r\theta'^{2}}{4} \right)A'-\frac{rA'^{3}\theta'^{2}}{16b^{2}H^{2}}&=& 0\nonumber\\
\theta'' +\theta' \left(\frac{4}{r}+\frac{F'}{F} \right)-\frac{rH\theta'^{3}}{4}-\frac{H^{2}}{r^{2}F}\frac{\partial V}{\partial \theta}&=&0\nonumber\\
\frac{H'}{H}-\frac{r}{4}\theta'^{2}&=&0\label{E8}
\end{eqnarray}
where the functions $ \mathcal{A} $, $ \mathcal{Q} $ and $ \mathcal{G} $ could be expressed as,
\begin{eqnarray}
\mathcal{A}&=&\frac{r \theta'^{2}}{12}-\frac{r^{2}\theta' \theta''}{6}-\frac{3}{r}\nonumber\\
\mathcal{Q}(r)&=& \frac{r\theta'^{2}H^{2}}{2}+\frac{H^{2}V(\theta)}{2r}-\frac{3H^{2}}{r}-\frac{rH^{2}\theta'^{2}}{12}V(\theta)-\frac{H^{2}\theta'}{6}\frac{\partial V}{\partial \theta}-\frac{A'^{2}}{12r}-\frac{A'A''}{6}\nonumber\\
\mathcal{G}(r,b^{2})&=&\frac{A'^{4}}{192b^{2}rH^{2}}-\frac{A'^{3}A''}{16b^{2}H^{2}}+\frac{rA'^{4}\theta'^{2}}{128b^{2}H^{2}}.
\end{eqnarray}

The next step would be to solve Eq.(\ref{E8}) perturbatively in the BI coupling ($ 1/b^{2} $). Since throughout this paper the computations have been performed keeping terms only upto leading order in the BI coupling ($ 1/b^{2} $), therefore we shall compute the corrected solutions only upto leading order in the perturbation series namely,
\begin{eqnarray}
F(r)&=&F^{(0)}(r)+\frac{1}{b^{2}}f(r)+\mathcal{O}(1/b^{4})\nonumber\\
H(r)&=&H^{(0)}(r)+\frac{1}{b^{2}}h(r)+\mathcal{O}(1/b^{4})\nonumber\\
A(r)&=&A^{(0)}(r)+\frac{1}{b^{2}}a(r)+\mathcal{O}(1/b^{4})\nonumber\\
\theta(r)&=&\theta^{(0)}(r)+\frac{1}{b^{2}}\vartheta(r)+\mathcal{O}(1/b^{4})\label{e9}
\end{eqnarray}
where $ f(r) $, $ h(r) $, $ a(r) $ and $ \vartheta(r) $ are the leading order BI corrections to $ F(r) $, $ H(r) $, $ A(r) $ and $ \theta(r) $ respectively.

Our next goal would be to substitute (\ref{e9}) in to (\ref{E8}) and identify equations order by order in the perturbation series. Let us first note down the equations corresponding to the zeroth order in the pertubative expansion (\ref{e9}). The equations corresponding to zeroth order in the BI coupling could be enumerated as follows,
\begin{eqnarray}
F'^{(0)}-\left(\frac{r \theta'^{(0)2}}{12}-\frac{r^{2}\theta'^{(0)} \theta''^{(0)}}{6}-\frac{3}{r} \right)F^{(0)}+\mathcal{Q}^{(0)}&=&0\nonumber\\
 A''^{(0)}+\left(\frac{2}{r}-\frac{r\theta'^{(0)2}}{4} \right)A'^{(0)}&=& 0\nonumber\\
 \theta''^{(0)} +\theta'^{(0)} \left(\frac{4}{r}+\frac{F'^{(0)}}{F^{(0)}} \right)-\frac{rH^{(0)}\theta'^{(0)3}}{4}-\frac{H^{(0)2}}{r^{2}F^{(0)}}\frac{\partial V}{\partial \theta^{(0)}}&=&0\nonumber\\
 \frac{H'^{(0)}}{H^{(0)}}-\frac{r}{4}\theta'^{(0)2}&=&0 
\end{eqnarray}
where $ \mathcal{Q}^{(0)} $ is the value of the function $ \mathcal{Q} $ when evaluated for the zeroth order solutions in the perturbation series. One interesting thing at this point is to note that the equations corresponding to $ F^{(0)} $ and $ H^{(0)} $ are easily solvable in the sense that these are basically linear first order differential equations whose solutions could be expressed as,
\begin{eqnarray}
F^{(0)}(r)&=&(I^{(0)}_{F})^{-1}\left[ -\int_{\infty}^{r}\ ds\ I^{(0)}_{F}(s)\mathcal{Q}^{(0)}(s) + \mathcal{C}_{1} \right] \nonumber\\
H^{(0)}(r)&=&exp\left[ {\int_{\infty}^{r} ds\ \frac{s}{4}\theta'^{(0)2}}\right] \label{F}
\end{eqnarray}
where,
\begin{eqnarray}
  I^{(0)}_{F}(r)=exp\left[ -\int_{\infty}^{r}\ ds\ \left(\frac{s \theta'^{(0)2}}{12}-\frac{s^{2}\theta'^{(0)} \theta''^{(0)}}{6}-\frac{3}{s} \right)\right]   
\end{eqnarray}
is the integrating factor and $\mathcal{C}_{1}$ is some arbitrary constant\footnote{By demanding the fact that $ F^{(0)}\sim 1 $ near the boundary of the $ AdS_{4} $ eventually fixes this constant uniquely.}.
Therefore the zeroth order solutions corresponding to the metric fluctuations are fully determined once we know the profile for the scalar as well as the gauge fields at zeroth order level. 

Our next aim would be to note down the solutions at the leading order in the BI coupling ($ 1/b^{2} $). In order to do that, in the following we enumerate all the equations corresponding to the leading order in the BI coupling namely, 
\begin{eqnarray}
f'-\left(\frac{r \theta'^{(0)2}}{12}-\frac{r^{2}\theta'^{(0)} \theta''^{(0)}}{6} \right)f + \mathcal{Q}^{(1)}&=&0\nonumber\\
a'' +\left(\frac{2}{r}-\frac{r\theta'^{(0)2}}{4} \right)a'+ \frac{A''^{(0)}A'^{2(0)}}{H^{2(0)}}-\frac{r}{2}\vartheta' A'^{(0)}-\frac{rA'^{3(0)}\theta'^{2(0)}}{16H^{2(0)}}&=&0\nonumber\\
\vartheta'' + \vartheta' \left(\frac{4}{r}+\frac{F'^{(0)}}{F^{(0)}}+\frac{3rH^{(0)}\theta'^{2(0)}}{4} \right)+\frac{\theta'^{(0)}}{F^{(0)}}\left(f'-\frac{F'^{(0)}f}{F^{(0)}} \right)- \mathcal{K} &=&0\nonumber\\
h'-\frac{H'^{(0)}}{H^{(0)}}h-\frac{rH^{(0)}\theta'^{(0)}\vartheta'}{2}&=&0\label{E9}
\end{eqnarray}
 where $ \mathcal{Q}^{(1)} $ and $ \mathcal{K} $ are some complicated functions of the radial coordinate ($ r $) whose exact expressions have been provided in the Appendix.
 
Like in the previous case here also we encounter two linear differential equations corresponding to the metric fluctuations and two second order differential equations  corresponding to the gauge as well as the matter field fluctuations at the leading order in the BI coupling ($ 1/b^{2} $). Like we did earlier, it is now quite straightforward to express the solutions corresponding to $ f(r) $ and $ h(r) $ which could be enumerated as follows,
\begin{eqnarray}
f(r)&=&(I_{f})^{-1}\left[ -\int_{\infty}^{r}\ ds\ I_{f}(s)\mathcal{Q}^{(1)}(s) + \mathcal{C}_{2} \right] \nonumber\\
h(r)&=&(I_{h})^{-1}\left[ \int_{\infty}^{r}\ ds\ I_{h}(s)\frac{sH^{(0)}(s)\theta'^{(0)}(s)\vartheta'(s)}{2} + \mathcal{C}_{3} \right]\label{f}
\end{eqnarray}  
where in the present case we have the following two integrating factors namely,
\begin{eqnarray}
I_f (r)&=&exp\left[ -\int_{\infty}^{r}\ ds\ \left(\frac{s \theta'^{(0)2}}{12}-\frac{s^{2}\theta'^{(0)} \theta''^{(0)}}{6}\right)\right]\nonumber\\ 
I_{h}(r)&=& exp\left[ -\int_{\infty}^{r}\ ds\ \frac{H'^{(0)}}{H^{(0)}} \right].
\end{eqnarray}

\subsection{Computation of $ \eta_{A} $}
Since the coefficient of hall viscosity ($ \eta_{A} $) could be uniquely determined from the knowledge of the spatial components of the stress tensor namely $ T_{xy} $ (or, $ T_{xx}-T_{yy} $), therefore for the present case it will be sufficient for us to solve $ \mathcal{O}(\varepsilon) $ equations in terms of the tensor modes with respect to the boundary spatial $ SO(2) $ symmetry.
Let us consider the $ xy $ component of the trace reversed Einstein's equation (\ref{E6}) namely,
\begin{eqnarray}
R^{(1)}_{xy}+3g^{(1)}_{xy}-\lambda C^{(1)}_{xy}=T_{xy}^{(1)}(\theta)+T_{xy}^{(1)}(F)\label{E10}
\end{eqnarray}
where the superscript $ (1) $ stands for the fluctuations corresponding to the leading order ($ \mathcal{O}(\varepsilon) $) in the derivative expansion. Computing each term in (\ref{E10}) we arrive at the following equation namely,
\begin{eqnarray}
\frac{1}{H(r,b^{2})}\frac{d}{dr}\left[-\frac{1}{2}\frac{r^{4}F(r,b^{2})}{H(r,b^{2})}\frac{d\alpha_{xy} }{dr}\right]+ \left[\frac{r^{3}H'F}{H^{3}}-\frac{r^{3}F'}{H^{2}}-\frac{3r^{2}F}{H^{2}}+3r^{2}-\frac{r^{2}}{2}V(\theta)-\frac{r^{2}A'^{2}}{4H^{2}}-\frac{3r^{2}A'^{4}}{64b^{2}H^{4}} \right]\alpha_{xy}\nonumber\\
= \frac{r}{H(r,b^{2})}(\partial_x \beta_{y}+\partial_y \beta_{x})+\frac{\lambda}{4H(r,b^{2})}\frac{d}{dr}\left[\frac{r^{4}F'(r,b^{2})\theta'(r,b^{2})}{H^{2}(r,b^{2})}\right](\partial_x \beta_{x}- \partial_y \beta_{y}).\nonumber\\   
\end{eqnarray}

Using (\ref{E7}) we finally obtain,
\begin{eqnarray}
\alpha_{xy}(r)=\int_{r}^{\infty}d\ell \frac{2H(\ell , b^{2})}{\ell^{4}F(\ell , b^{2})}\int_{r_h}^{\ell} ds\left[ s(\partial_x \beta_{y}+\partial_y \beta_{x})+\frac{\lambda}{4}\frac{d}{ds}\left(\frac{s^{4}F'(s,b^{2})\theta'(s,b^{2})}{H^{2}(s,b^{2})}\right)(\partial_x \beta_{x}- \partial_y \beta_{y})\right].\label{E11}
\end{eqnarray}

From (\ref{E11}) it is quite evident that one can in fact simplify this expression a bit further in order to get an exact analytical expression for $ \alpha_{xy} $ upto leading order in the BI coupling ($ 1/b^{2} $). However, we are not going to do any such truncation at this stage and will do it in future.  

Note that for an asymptotically $ AdS $ space time the boundary stress tensor for odd boundary dimensions could be formally expressed as \cite{Chen:2011fs},
\begin{eqnarray}
\langle T_{ij}\rangle=\frac{d}{16\pi G_N}g_{(d)ij}\label{E12}
\end{eqnarray} 
where $ g_{(d)ij} $s are the coefficients of the following metric expansion namely,
\begin{eqnarray}
g(x^{a},r)= g_{(0)}+\frac{1}{r^{2}}g_{(2)}+..~~..+\frac{1}{r^{d}}g_{(d)}+..~~..\label{E13}
\end{eqnarray}
Moreover in the limit $ r\rightarrow\infty $ one can in fact show that,
\begin{eqnarray}
r^{m}\alpha_{xy}(r)=-\frac{r^{m+1}}{m}\frac{d\alpha_{xy}}{dr}.\label{E14}
\end{eqnarray} 

Using (\ref{E12}), (\ref{E13}) and (\ref{E14}) the boundary stress tensor eventually turns out to be,
\begin{eqnarray}
\langle T_{xy}\rangle=\frac{3}{16\pi G_N}\alpha_{(3)xy}\label{T}
\end{eqnarray} 
which means that all we need to do is to identify the constant part of the entity $ r^{3}\alpha_{xy} $.

 In order to proceed let us first note that,
\begin{eqnarray}
r^{3}\alpha_{xy}=-\frac{r^{4}}{3}\frac{d}{dr}\int_{r}^{\infty}d\ell \frac{2H(\ell , b^{2})}{\ell^{4}F(\ell , b^{2})}~~ ~~ ~~ ~~ ~~ ~~ ~~ ~~ ~~ ~~ ~~ ~~ ~~ ~~ ~~ ~~ ~~ ~~ ~~ ~~ ~~ ~~ ~~ ~~ ~~ ~~ ~~ ~~ ~~ ~~ ~~ ~~ ~~
\nonumber\\  \times \int_{r_h}^{\ell}
 ds\left[ s(\partial_x \beta_{y}+\partial_y \beta_{x})+\frac{\lambda}{4}\frac{d}{ds}\left(\frac{s^{4}F'(s,b^{2})\theta'(s,b^{2})}{H^{2}(s,b^{2})}\right)(\partial_x \beta_{x}- \partial_y \beta_{y})\right].\label{E15}
\end{eqnarray}

In order to evaluate (\ref{E15}) one first needs to evaluate the ratio $ \frac{H(r,b^{2})}{F(r,b^{2})} $ near the boundary of the $ AdS_{4} $ upto leading order in the BI coupling ($ 1/b^{2} $). This is indeed a non trivial check as the metric we are considering is fully back reacted. The ratio is given by,
\begin{eqnarray}
\frac{H(r,b^{2})}{F(r,b^{2})}=\frac{H^{(0)}}{F^{(0)}}\left[ 1+\frac{1}{b^{2}}\left(\frac{h}{H^{(0)}}-\frac{f}{F^{(0)}} \right) \right]+\mathcal{O}(1/b^{4}).
\end{eqnarray}

Using (\ref{F}) and (\ref{f}) and taking $ m^{2}=-2 $ and thereby considering $ \theta(r) \sim \frac{\langle\mathcal{O}\rangle}{r^{2}} $ and $ A\sim \mu -\frac{\varrho}{r} $ near the boundary of the $ AdS_{4} $ we note that\footnote{The above boundary behaviour of $ \theta $ eventually corresponds to a spontaneous breaking of parity in the boundary CFT \cite{Chen:2011fs},\cite{Chen:2012ti}.},
\begin{eqnarray}
\frac{H^{(0)}}{F^{(0)}}&=&1+\mathcal{O}(1/r^{2})\nonumber\\
\frac{h}{H^{(0)}}&=&1+\mathcal{O}(1/r^{4})\nonumber\\
\frac{f}{F^{(0)}}&=&1+\mathcal{O}(1/r^{2}).
\end{eqnarray}

This eventually suggests that
\begin{eqnarray}
\frac{H(r,b^{2})}{F(r,b^{2})}|_{r\rightarrow\infty}=1+\mathcal{O}(1/r^{2}).\label{E16}
\end{eqnarray}

This is indeed an important observation in the sense that even in the presence of higher derivative terms in the Maxwell sector of the theory, at the leading order in the BI coupling we do not get any correction to the above ratio (\ref{E16}) near the boundary of the $AdS_4$. This has an important consequence on the earlier observation that the coefficient of shear viscosity ($ \eta_{s} $) does not get modified due to BI corrections at the leading order \cite{Cai:2008in}. From the analysis based on the fluid gravity approach we can see that this happens due to the mutual cancellation of the finite pieces appearing from the metric near the boundary of the $ AdS_4 $.

Finally using the above relation in (\ref{E16}) and substituting (\ref{E15}) in to (\ref{T}) we note that,
\begin{eqnarray}
\langle T_{xy}\rangle_{r\rightarrow\infty} = \eta_{s}(\partial_x \beta_{y}+\partial_y \beta_{x})|_{r=r_{h}}+\eta_{A}(\partial_x \beta_{x}- \partial_y \beta_{y})|_{r=r_{h}}
\end{eqnarray}
where,
\begin{eqnarray}
\eta_{s}= \frac{1}{16\pi G_{N}}
\end{eqnarray}
is the usual shear viscosity mode that does not get corrected at the leading order in the BI coupling ($ 1/b^{2} $) due to the fact as mentioned above. On the other hand, the coefficient of hall viscosity to entropy ratio turns out to be,
\begin{eqnarray}
\frac{\eta_{A}}{s}&=&\frac{\eta^{(0)}_{A}}{s}\left[ 1+\frac{1}{b^{2}}\Xi (M,Q)\right] +\mathcal{O}(1/b^{4}).\label{E17}
 \end{eqnarray} 
where the individual terms could be enumerated as,
\begin{eqnarray}
\frac{\eta^{(0)}_{A}}{s}&=&-\frac{\lambda}{8\pi}\left[\frac{r^{4}F'^{(0)}(r)\theta'^{(0)}(r)}{H^{2(0)}(r)}
 \right]_{r=r_{h}}\nonumber\\
 \Xi (M,Q)&=&\left[ \frac{f'}{F'^{(0)}}+\frac{\vartheta'}{\theta'^{(0)}}-\frac{2h}{H^{(0)}}\right]_{r=r_{h}}.\label{E18}
\end{eqnarray}
 
 The first term on thr r.h.s. of (\ref{E17}) is the hall viscosity coefficient  corresponding to the Einstein Maxwell theory of gravity in $ AdS_{4} $ in the presence of the back reaction. On the other hand, the second term ($ \Xi $) is the non trivial correction appearing at the leading order in the BI coupling. The correction ($ \Xi $) appearing in (\ref{E18}) is the most general correction in the presence of back reaction. 
 
 \subsection{A sample calculation in the probe limit}
In order to get some intuitive idea regarding the value of this leading order correction coefficient ($\Xi $) let us imagine a simpler situation where we consider the charged black brane solution in the \textit{probe} limit namely,
 \begin{eqnarray}
 ds^{2}=2dv dr - r^{2}F(r)dv^{2}+r^{2}(dx^{2}+dy^{2})\label{E19}
 \end{eqnarray}
where,
\begin{eqnarray}
F(r)=1-\frac{2M}{r^{3}}+\frac{Q^{2}}{r^{4}}-\frac{Q^{4}}{20b^{2}r^{8}}+\mathcal{O}(1/b^{4})\label{E20}
\end{eqnarray} 
 is the non linear generalization of the charged black brane solution for the usual Einstein Maxwell theory in ($ 3+1 $) dimensions \cite{Chen:2011fs}. From the above solutions (\ref{E19}) and (\ref{E20}) we note that,
 \begin{eqnarray}
 F^{(0)}(r)&=&1-\frac{2M}{r^{3}}+\frac{Q^{2}}{r^{4}}\nonumber\\
 f(r)&=& -\frac{Q^{4}}{20r^{8}}\label{e2}
\end{eqnarray}  
 along with $ H^{(0)}=1 $ and $ h=0 $. Next we would like to solve the fluctuations for the axionic field ($ \theta $) considering the special case $ V(\theta)=0$. In the following we enumerate the set of equations as,
 \begin{eqnarray}
 \theta''^{(0)}++\theta'^{(0)} \left(\frac{4}{r}+\frac{F'^{(0)}}{F^{(0)}} \right)&=&0\nonumber\\
 \vartheta'' +\frac{F'^{(0)}}{F^{(0)}}\vartheta' +\frac{\theta'^{(0)}}{F^{(0)}}\left(f' -\frac{fF'^{(0)}}{F^{(0)}} \right)&=&0.\label{E21} 
 \end{eqnarray}
 
 Integrating once the above set of equations (\ref{E21}) we note that,
 \begin{eqnarray}
 \theta'^{(0)}(r)&=&\frac{\mathcal{G}}{r^{4}-2 Mr +Q^2}\approx \frac{\mathcal{G}}{r^{4}}\left( 1+\frac{2M}{r^{3}}-\frac{Q^{2}}{r^{4}}\right) + ..~..\nonumber\\
\vartheta'(r)&=& \mathcal{N} \exp \left(-2 \left(-\frac{M^2}{\text{r}^6}+\frac{M Q^2}{\text{r}^7}-\frac{M}{\text{r}^3}-\frac{Q^4}{4 \text{r}^8}+\frac{Q^2}{2 \text{r}^4}\right)\right)(1+ \mathcal{Z}(M,Q,r))
 \label{NE22} 
 \end{eqnarray}
 where $ \mathcal{G} $ and $ \mathcal{N} $ are two arbitrary constants and $ \mathcal{Z} $ is given by,
 \begin{eqnarray}
 \mathcal{Z}(M,Q,r)=-\int_1^{\text{r}} \frac{2 Q^4 \exp \left(2 \left(-\frac{M^2}{s^6}+\frac{M Q^2}{s^7}-\frac{M}{s^3}-\frac{Q^4}{4 s^8}+\frac{Q^2}{2s^4}\right)\right)}{5 s^8} \, ds.
 \end{eqnarray}
 
 In order to make the computation a bit simpler we exploit the scaling symmetry\footnote{By scaling symmetry we mean the transformations $ r\rightarrow ar $, $ (v,x,y)\rightarrow a^{-1}(v,x,y) $, $ M\rightarrow a^{3}M $, $ Q\rightarrow a^{2}Q $ under which the metric (\ref{E19}) remains invariant.} of the metric (\ref{E19}) to rescale $ r=r_{h}=1 $. Finally, using (\ref{e2}) and (\ref{NE22}), from (\ref{E18}) we obtain\footnote{Where we have set $ \mathcal{N}=1 $ and $ \mathcal{G}=1 $.},
\begin{eqnarray}
 \Xi (M,Q)&=&\frac{Q^{4}}{5(3M-2Q^{2})}+(1-2 M +Q^2)\exp \left(-2 \left(-M^{2}+M Q^{2}-M-\frac{Q^4}{4}+\frac{Q^2}{2}\right)\right).\nonumber\\
\end{eqnarray}
 
This observation is interesting in the sense that here we are having some finite corrections to the $ \eta_{A}/s $ ratio due to the higher derivative ($\sim  F^{4} $) corrections appearing in the abelian gauge sector of the theory. This situation is completely reverse what people had found earlier while studying the shear viscosity to entropy ratio in the presence of BI corrections to the usual Maxwell action \cite{Cai:2008in}. There it was observed that the shear viscosity to entropy ratio ($ \eta_{s}/s $) is not at all affected due to such corrections and therefore it was concluded that the violation to the universal bound on the $ \eta_{s}/s $ ratio is caused solely due to the the higher derivative corrections on the gravity side \cite{Cai:2008in},\cite{Tan:2009yg}.

\section{ $ R^{2} $ gravity}

In the previous section we have discussed the effect of adding higher derivative ($\sim  F^{4} $) corrections of the gauge fields on the coefficient of hall viscosity ($ \eta_{A} $). In this section our goal to is to study the effect of adding $ R^{2} $ corrections on the $ \eta_{A}/s $ ratio, where by $ R^{2} $ corrections we mean the higher derivative corrections to the Einstein gravity in ($ 3+1 $) dimensions. The action that we consider as the starting point of our analysis is the following,
\begin{eqnarray}
S=\int d^{4}x\sqrt{-g}\left[R+\alpha R^{2}+\beta R^{2}_{ab}- 2\Lambda - \frac{1}{2}(\partial \theta)^{2}- V(\theta)- \frac{\lambda \theta}{8} \frac{\epsilon^{abcd}}{\sqrt{-g}}R^{m}\ _{ncd}R^{n}\ _{mab} \right] 
\end{eqnarray} 
where $ \alpha $ and $ \beta $ are two coupling constants. 

Our aim in this section is to study the effect of these higher curvature corrections on the coefficient of hall viscosity ($ \eta_{A} $) at the leading order in the coefficients $ \alpha $ and $ \beta $. In order to do that let us first consider the following redefinition \cite{Brigante:2007nu} of the metric namely,
\begin{eqnarray}
g_{ab}\rightarrow \tilde{g}_{ab}=g_{ab} +\tilde{\beta} R_{ab}+ \tilde{\alpha} g_{ab}R \label{E22}
\end{eqnarray}
where $ \tilde{\beta} $ and $\tilde{\alpha}  $ are another set of coupling constants.

 Under the above redefinition (\ref{E22}) of the field ($ g_{ab} $) the action changes as,
\begin{eqnarray}
S \rightarrow \tilde{S} = S+\delta S.\label{E23}
\end{eqnarray}

Our next goal is to compute the second term on the r.h.s of (\ref{E23}) upto sixth derivative order in the metric ($ g_{ab} $). Let us split $ \delta S $ into various components namely\footnote{We drop out all the total derivative terms.},
\begin{eqnarray}
(\delta S)_{I}&=&=\sqrt{-g}\left( \frac{\tilde{\beta}}{2}+3\tilde{\alpha}\right)  R^{2}+\sqrt{-g}\tilde{\beta} R^{ab}R_{ab}\label{NE24}\\
(\delta S)_{II}&=&\alpha \sqrt{-g}\left(\frac{\tilde{\beta}}{2}+4\tilde{\alpha} \right)R^{3}+2\sqrt{-g}\alpha\tilde{\beta} R R^{ab}R_{ab} 
+ 2\sqrt{-g}\alpha \left( \tilde{\beta} R^{ab}\nabla_a \nabla_b R 
-(\tilde{\beta} +3\tilde{\alpha}) R \nabla_{a}^{2}R\right)\nonumber\\  
\end{eqnarray} 
\begin{eqnarray}
(\delta S)_{III}=\sqrt{-g}\beta\left(\frac{\tilde{\beta}}{2}+2\tilde{\alpha} \right)R R^{ab}R_{ab}-\sqrt{-g}\beta \tilde{\beta} R^{ab}\nabla_{a}\nabla_{b}R~~~~~~~~~~~~\nonumber\\
-2\tilde{\alpha} \beta\sqrt{-g} R^{ab} \nabla_a \nabla_b R -\sqrt{-g}\beta \tilde{\beta} R_{ab}\nabla_m^{2}R^{ab}-\tilde{\alpha}\beta\sqrt{-g} R \nabla_a^{2}R + 2\sqrt{-g}\beta \tilde{\beta} \nabla_a \nabla_m R^{ab}R^{m}\ _{b}   
\end{eqnarray}
\begin{eqnarray}
(\delta S)_{IV}&=& \frac{\sqrt{-g}R}{2}\left(\tilde{\beta} +4 \tilde{\alpha} \right) \left(- 2\Lambda - \frac{1}{2}(\partial \theta)^{2}- V(\theta) \right)\\
(\delta S)_{V}&=& -\lambda C^{ab}(\tilde{\beta} R_{ab}+\tilde{\alpha} g_{ab}R)=-\lambda\tilde{\beta} C^{ab}R_{ab}\label{E25}
\end{eqnarray}
where $ C_{ab} $ is the symmetric (traceless) Cotton tensor that has been defined earlier in (\ref{e6}).

 A number of things are to be noted at this stage. First of all from the above set of equations (\ref{NE24})-(\ref{E25}) we note that the entities like $ (\delta S)_{I} $ and $ (\delta S)_{IV} $ that appear at the leading order in the couplings namely $ \alpha $ and $ \beta $ are essentially the terms which contain derivatives upto quartic order in the metric ($ \sim (\partial^{4} g_{ab}) $). On the other hand the remaining entities like $ (\delta S)_{II} $, $ (\delta S)_{III} $ and $ (\delta S)_{V} $ contain sixth order derivatives of the metric ($ \sim (\partial^{6} g_{ab}) $). Therefore to start with if we retain terms upto quartic derivative order in the action, then we may simply drop the terms like $ (\delta S)_{II} $, $ (\delta S)_{III} $ and $ (\delta S)_{V} $ as they appear in the higher order derivatives of the metric ($ g_{ab} $). In that case we are simply left with,
 \begin{eqnarray}
 \tilde{S}= S_{H}+ S^{(1)}\label{S}
\end{eqnarray}     
 where the two terms on the r.h.s of (\ref{S}) could be expressed as,
 \begin{eqnarray}
 S_H&=&\int d^{4}x\sqrt{-g}\left[R- 2\Lambda - \frac{1}{2}(\partial \theta)^{2}- V(\theta)- \frac{\lambda \theta}{8} \frac{\epsilon^{abcd}}{\sqrt{-g}}R^{m}\ _{ncd}R^{n}\ _{mab} \right]
 \end{eqnarray}
 and,
 \begin{eqnarray}
 S^{(1)}=\left(\alpha + 3 \tilde{\alpha} +\frac{\tilde{\beta}}{2} \right)\int d^{4}x\sqrt{-g}R^{2}+(\beta + \tilde{\beta})\int d^{4}x\sqrt{-g}R_{ab}R^{ab}\nonumber\\
 -\left(2\tilde{\alpha}+\frac{\tilde{\beta}}{2} \right)\int d^{4}x\sqrt{-g} R \left( 2\Lambda + \frac{1}{2}(\partial \theta)^{2}+V(\theta) \right)
 \end{eqnarray}
 respectively.
 
 One can now set $ S^{(1)} $ to be equal to zero unambiguously imposing the following constraints on the coupling constants namely,
 \begin{eqnarray}
 \alpha + 3 \tilde{\alpha} +\frac{\tilde{\beta}}{2} &=&0\nonumber\\
 \beta + \tilde{\beta}&=&0\nonumber\\
 2\tilde{\alpha}+\frac{\tilde{\beta}}{2}&=&0
\end{eqnarray}  
 which eventually determines three of the constants ($ \beta  $, $ \tilde{\beta} $ and $  \tilde{\alpha}$) in terms of one coupling constant ($ \alpha $) namely,
 \begin{eqnarray}
 \beta =-4\alpha , ~~~\tilde{\beta}= 4\alpha ,~~~\tilde{\alpha} =-\alpha.\label{c}
 \end{eqnarray}
 
 Therefore with the above choice (\ref{c}) of the coupling constants, we are finally left with $ S_H $ (which is the action originally proposed in \cite{Saremi:2011ab}) and the action $ S^{(2)} $ at the sixth derivative level namely,
 \begin{eqnarray}
S^{(2)} =2\alpha^{2}\int\sqrt{-g}\left[8R_{ab}\nabla^{2}_{m}R^{ab}+8R^{ab}\nabla_a \nabla_b R -16\nabla_a \nabla_m R^{ab}R^{m}\ _{b}-3R\nabla_a^{2}R-4RR_{ab}^{2}-R^{3} \right]+\mathcal{S}
 \end{eqnarray}
 where,
 \begin{eqnarray}
 \mathcal{S}=-4\lambda\alpha\int d^{4}x\sqrt{-g}C^{ab}R_{ab}.
 \end{eqnarray}
 
 Therefore the $ \eta_{A}/s $ ratio at the quartic derivative level could be expressed as\footnote{Once again the superscript ($0$) stands for the unperturbed solutions.},
 \begin{eqnarray}
 \frac{\eta_{A}}{s}=-\frac{\lambda}{8\pi}\left[\frac{r^{4}F'^{(0)}(r)\theta'^{(0)}(r)}{H^{2(0)}(r)}\right]_{r=r_h} +\mathcal{O}(\alpha^{2}).
\end{eqnarray}  
 
  From the above analysis we conclude that the higher derivative $ R^{2} $ corrections do not affect the $ \eta_{A}/s $ ratio at the \textit{leading} order in the couplings $ \alpha $. Also form the above analysis we note that if we consider higher derivative corrections beyond quartic order then there is a finite probability for $ \eta_{A}/s $ ratio to be corrected at the sixth derivative level.

\section{Summary and final remarks}
Let us now summarize what we have done so far. The objective of the present analysis was to explore the hall viscosity to entropy ratio ($\eta_{A}/s  $) in presence of higher derivative corrections both on the gauge sector as well as on the gravity sector of the original theory proposed by Saremi and Son in \cite{Saremi:2011ab}. In order to address this issue in the first part of our analysis we note that the higher derivative ($ \sim F^{4} $) corrections imposed on the gauge sector of the theory modifies the $\eta_{A}/s  $ ratio at the leading order in the coupling constant. This result therefore illustrates a somewhat different scenario which shows that the higher derivative corrections on the gauge sector could modify the $\eta_{A}/s  $ ratio while on the other hand it does not affect the $\eta_{s}/s  $ ratio \cite{Cai:2008in},\cite{Tan:2009yg}.
In the second part of our analysis we consider $ R^{2} $ corrections on the Einstein Hilbert sector of the theory \cite{Saremi:2011ab} where we note that $\eta_{A}/s  $ ratio does not receive any correction upto the quartic derivative order. On the other hand our calculations show that the ratio might get corrected at the sixth derivative level which we leave as an interesting exercise to be pursed in future. Finally it would also be an interesting exercise to carry out similar analysis in presence of other non trivial matter couplings in four dimensions. Also it would be an interesting project to explore the effect of various other parity odd terms (like  $ R\wedge F\wedge F $) on the $\eta_{A}/s  $ ratio.

{\bf {Acknowledgements :}}

It is a great pleasure to thank Aninda Sinha, Sayantani Bhattacharyya and Nabamita Banerjee for their valuable comments. The author also would like to acknowledge the financial support from CHEP, Indian Institute of Science, Bangalore.\\

\appendix

\noindent
{\bf \large Appendix}\\
The exact analytical expressions for the radial functions appearing in (\ref{E9}) are the following,
\begin{eqnarray*}
 \mathcal{Q}^{(1)}=\frac{r^{2}}{6}\left(\theta'^{(0)}\vartheta'' +\vartheta' \theta''^{(0)}-\frac{\theta'^{(0)}\vartheta'}{r} \right)F^{(0)}+ r\theta'^{(0)2}H^{(0)}h+r\vartheta' \theta'^{(0)}H^{2(0)}
 + \frac{m^{2}H^{(0)}\theta^{(0)}}{2r}\left( H^{(0)}\vartheta +h\theta^{(0)}\right)~~ ~~ ~~ ~~ ~~ ~~ ~~ ~~ ~~ ~~\nonumber\\
-\frac{A'^{(0)}a'}{6r}
+\frac{cH^{(0)}\theta^{(0)4}}{2r} \left(\frac{H^{(0)}\vartheta}{\theta^{(0)}}+\frac{h}{2} \right)-\frac{6}{r}H^{(0)}h
-\frac{m^{2}r}{12}\theta^{(0)}\theta'^{(0)}H^{(0)}\left(H^{(0)}\theta'^{(0)}\vartheta +H^{(0)}\vartheta' \theta^{(0)}+h\theta'^{(0)}\theta^{(0)} \right)~~ ~~ ~~ ~~ ~~ ~~ ~~ ~~ ~~ ~~\nonumber\\
-\frac{1}{6} \left( A'^{(0)}a'' + a'A''^{(0)}\right)-\frac{rc}{12}\theta'^{(0)}\theta^{4(0)}H^{(0)} \left( \frac{H^{(0)}\theta'^{(0)}\vartheta}{\theta^{(0)}}+\frac{H^{(0)}\vartheta'}{2}+\frac{h}{2}\right)-\frac{m^{2}H^{(0)2}}{6}\left( \theta'^{(0)}\vartheta +\vartheta' \theta^{(0)}\right)~~ ~~ ~~ ~~ ~~ ~~ ~~ ~~ ~~ ~~ ~~ \nonumber\\
-\frac{m^{2}H^{(0)2}\theta^{(0)}h\theta'^{(0)}}{3}-\frac{cH^{(0)}\theta^{(0)3}}{6}\left( \frac{3H^{(0)}\vartheta}{\theta^{(0)}}+H^{(0)}\vartheta' + 2h\theta''^{(0)}\right)-\frac{A'^{3(0)}}{16H^{2(0)}}\left(A''^{(0)}-\frac{A'^{(0)}}{12r} -\frac{rA'^{(0)}\theta'^{2(0)}}{8}\right)~~ ~~ ~~ ~~ ~~ ~~ ~~ ~~ ~~ ~~ 
 \end{eqnarray*}
 and,
\begin{eqnarray*}
\mathcal{K}=\frac{rh\theta'^{3(0)}}{4}+\frac{m^{2}H^{(0)}}{r^{2}F^{(0)}} \left(H^{(0)}\vartheta -\frac{\theta^{(0)}H^{(0)}f}{F^{(0)}}+2h\theta^{(0)} \right)+\frac{cH^{(0)}\theta'^{3(0)}}{r^{2}F^{(0)}}\left(\frac{3\vartheta H^{(0)}}{\theta^{(0)}}-\frac{H^{(0)}f}{F^{(0)}}+2h \right).  
\end{eqnarray*}


\end{document}